\magnification=\magstep1
\baselineskip=15pt
%
%
%
%%%%%   M A C R O S   A N D   R E F E R E N C E S  %%%%%%%
%%%%%%%%%%%%%%%%%%%%%%%%%%%%%%%%%%%%%%%%%%%%%%%%%%%%%%%%%%
%
%
%
%   ENUMERATION MACROS
%   ==================
%
% CHAPTER/SECTION ENUMERATION
%
%
%

\def\Numerujstart#1{\count21=#1
\global\advance\count21 by -1
}

\def\Numeruj{%
\global\advance\count21 by 1
%\leavevmode\hbox{\number\count21}%
\xdef\NumerujNO{\number\count21}
}
\Numerujstart 1

\def\Sectionstart#1{\count12=#1
\global\advance\count12 by -1
}

\def\Section{%
\global\advance\count12 by 1
\leavevmode\hbox{\number\count12}%
\xdef\SecNO{\number\count12}%
\Numerujstart1%
}
\Sectionstart 1

\def\Chapterstart#1{\count11=#1
\global\advance\count11 by -1
}

\def\Chapter{%
% Resetting the EQNOD, Section counters:
\Sectionstart 1
\EQNODstart 1
\global\advance\count11 by 1
\leavevmode\hbox{\number\count11}%
\xdef\ChapNO{\number\count11}%
}
\Chapterstart 1

% EQUATION ENUMERATION

\def\EQNOstart#1{\count13=#1
\global\advance\count13 by -1 }

\def\EQNO#1{%
\global\advance\count13 by 1
\eqno(\number\count13)%
\xdef#1{(\number\count13)}
}

\def\aEQNO#1{%
\global\advance\count13 by 1
(\number\count13)%
\xdef#1{(\number\count13)}
}

\EQNOstart 1

\def\EQNODstart#1{\count14=#1
\global\advance\count14 by -1    }

\def\EQNOD#1{%
\global\advance\count14 by 1
\eqno(\ChapNO.\number\count14)%
\xdef#1{(\ChapNO.\number\count14)}
}

\def\aEQNOD#1{%
\global\advance\count14 by 1
(\ChapNO.\number\count14)%
\xdef#1{(\ChapNO.\number\count14)}
}

\EQNODstart 1

% REFERENCES ENUMERATION

\def\REFstart#1{\count15=#1
\global\advance\count15 by -1    }

\def\REF#1{%
\global\advance\count15 by 1
%%%   Ref. Type defined here:
\number\count15%
\xdef#1{\number\count15}%
}

\def\REFpr#1{%
\global\advance\count15 by 1%
%%%   Ref. Type defined here:
$^{\number\count15}$%
\xdef#1{$^{\number\count15}$}%
}
\REFstart 1

\def\RIEMANNref{
G.F.B.~Riemann, in {\it \"Uber die Hypothesen, welche der
Geometrie zugrunde liegen},
edited by H.~Weyl (Springer Verlag, Berlin 1919). }

\def\AMBARCUMIANref{
V.~Ambarcumian, D.~Ivanienko, {\it Z.~Phys.} {\bf 64} (1930) 563. }

\def\DIRACQUANTref{
P.A.M.~Dirac, {\it Proc.~Roy.~Soc.} {\bf A109} (1926) 642,
Sec.~3.  }

\def\WILSONref{
K.~Wilson, {\it Phys.~Rev.} {\bf D10} (1974) 2445.   }

\def\REGGEref{
T.~Regge, {\it Nuovo Cimento} {\bf 19} (1961) 558.   }

\def\GIBBSref{
P.~Gibbs, {\it The Small Scale Structure of Space--Time}
preprint PEG--06--95 (1995) hep--th/9506171. }

\def\BEKENSTEINref{
J.D.~Bekenstein, {\it Lett.~Nuovo Cimento} {\bf 11} (1974) 467.   }

\def\BEKENSTEINMUKHANOVref{
J.D.~Bekenstein, {\it Phys.~Lett.} {\bf B360} (1995) 7.   }

\def\MUKHANOVref{
V.F.~Mukhanov, {\it JETP Lett.} {\bf 44} (1986)
63.   }

\def\JACOBSONref{
T.~Jacobson, {\it Phys.~Rev.} {\bf D44} (1991) 1731.   }

\def\AHARONOVPETERSENref{
Y.~Aharonov, H.~Pendelton, A.~Petersen, {\it Int.~J.~Theor.~Phys.}
{\bf 2} (1969) 213.
}

\def\PENROSEspincalcref{
R.~Penrose, in: {\it Quantum theory and beyond},
ed. E.A.~Bastin (Cambridge University Press 1971).    }

\def\KLEBANOVref{
I.~Klebanov, L.~Suskind, {\it Nucl.~Phys.} {\bf B309} (1988)
175.   }

\def\LOOPREPRref{
G.~Horowitz, talk at the Conference on Quantum Gravity, Durham
(1994).  }

\def\COLEref{
E.~A.~B.~Cole, {\it Nuovo Cimento} {\bf 66A} (1970) 645. }

\def\TURBINERSMIRref{
Yu.~Smirnov, A.~Turbiner, {\it Mod.~Phys.~Lett.}
{\bf 10} (1995) 1795. }

\def\LUKIERPoincareref{
J.~Lukierski, A.~Nowicki, H.~Ruegg, {\it Phys.~Lett.} {\bf B293}
(1992) 344.  }

\def\FRAPPATref{
L.~Frappat, A.~Sciarrino, {\it Phys.~Lett.} {\bf B347}
(1995) 28.  }

\def\PABLOPRLref{
P.O.~Mazur, {\it Phys.~Rev.~Lett.} {\bf 57} (1986) 929;
{\bf 59} (1987) 2380.   }

\def\PABLOAPParef{
P.O.~Mazur,  {\it Acta Phys.~Pol.} {\bf B26} (1995) 1685.   }

\def\PABLOAPPbref{
P.O.~Mazur, {\it Acta Phys.~Pol.} {\bf B27} (1996) 1849.   } .

\def\GARAYref{
L.J.~Garay, {\it Int.~J.~Mod.~Phys.} {\bf A10} (1995) 145.   }

\def\THOOFTCAref{
G.~'t Hooft, {\it Nucl.~Phys.} {\bf B342} (1990)
471.       }

\def\THOOFTaref{
G.~'t Hooft, {\it Class. Quantum Grav.} {\bf 10} (1993)
1023.       }

\def\THOOFTbref{
G.~'t Hooft, {\it Class. Quantum Grav.} {\bf 10} (1993)
1653.               }

\def\Zygmuntref{A. Zygmunt, {\it Trigonometric Series, Cambridge University Press}, 1959.}

\def\WIDOMref{H. Widom, in {\it Studies in real and complex analysis}, Vol.3,
I.I. Hirschman, Jr. editor.}

\def\SMOLINref{
C.~Rovelli, L.~Smolin, {\it Nucl.~Phys.} {\bf B442} (1995) 593.    }

\def\SORKINref{
L.~Bombelli, J.~Lee, D.~Meyer, R.~D.~Sorkin,
{\it Phys.~Rev.~Lett.} {\bf 59} (1987) 521.
}

\def\DORFMEISTERref{G.~Dorfmeister, J.~Dorfmeister,
{\it J.~Func.~Anal.} {\bf 57} (1984) 301.
}

\def\COUTINHOref{S.~C. ~Coutinho, {\it A primer of Algebraic D-modules,
LMSST 33, Cambridge Universtiy Press}, 1995.
}

\def\DOUGLASref{M.R.~Douglas,
{\it Phys.~Lett.} {\bf B238} (1990) 176.
}

%
%
%  SPECIAL MACROS FOR THIS PAPER BY YACEK :
%  ----------------------------------------
%
\def\Box{\lower0.40ex\hbox{\lasy\kern0.05em\char"32\kern0.05em}}

\def\Proof{{\rm\bf Proof}}
\font\tenmsb=msbm10
\font\sevenmsb=msbm6
\font\fivemsb=msbm5
\newfam\msbfam
\textfont\msbfam=\tenmsb
\scriptfont\msbfam=\sevenmsb
\scriptscriptfont\msbfam=\fivemsb
\def\Bbb#1{\fam\msbfam\relax#1}
\font\teneufm=eufm10
\font\seveneufm=eufm7
\font\fiveeufm=eufm5
\newfam\eufmfam
\textfont\eufmfam=\teneufm
\scriptfont\eufmfam=\seveneufm
\scriptscriptfont\eufmfam=\fiveeufm

% small fonts
\font\eightrm=cmr8
\font\eightbf=cmbx8
\font\eightit=cmti8
\font\eightsl=cmsl8
\font\eightmus=cmmi8
\def\smalltype{\let\rm=\eightrm \let\bf=\eightbf \let\it=\eightit
\let\sl=\eightsl \let\mus=\eightmus \baselineskip=9.5pt minus .75pt
\rm}

%
%

%  MY SPECIAL MACROS FOR THIS PAPER:
%
\font\titlefont=cmbx12 scaled \magstep1
\def\at{@}

     \font\lasy=lasy10 scaled 1440
\def\Box{\lower0.40ex\hbox{\lasy\kern0.05em\char"32\kern0.05em}}
     \font\msam=msam10

\def\blackbox{\lower0.00ex\hbox{\msam\kern0.00em\char"4\kern0.05em}}

\def\sign{\hbox{\rm sign}}

\def\R{{\Bbb R}}  % real, complex & integer sets
\def\C{{\Bbb C}}
\def\Z{{\Bbb Z}}

\def\H{H}    % Hilbert space

%%% REFERENCES ARE INCLUDED HERE:
%%%

\def\REEDSIMONref{M.~Reed, B.~Simon, {\it Methods of Modern Mathematical
Physics,} Vols. 1 and 2 (Academic Press NY 1975). }

\hskip0.6\hsize Univ. of South Carolina

\hskip0.6\hsize preprint TH/9703--1

\vskip1.5\baselineskip

{\titlefont
\centerline{On Pairs of Difference Operators}

\vskip0.5\baselineskip
\centerline{Satisfying: [D,X] = Id}
}

\vskip3\baselineskip

\centerline{{\bf Andrzej Z. G\'orski}$\,$\footnote{$^{\dagger}$}{The
Ko\'sciuszko Foundation Fellow. Permanent
address: Institute of Nuclear Physics, Radzikowskiego 152,
31--342 Krak\'ow, Poland.}$\, ^1$
$\,${\bf and}$\,$ {\bf Jacek Szmigielski}$\, ^2$}

\vskip\baselineskip

\leftskip=0.5cm   \rightskip=1.0cm

\item{$^1$}{\it Department of Physics and Astronomy,
University of South Carolina,
\item{} Columbia, SC 29208, U.S.A.,
E--mail: {\tt gorski\at alf.ifj.edu.pl}}

\item{$^2$}{\it Department of Mathematics and Statistics,
University of Saskatchewan,
\item{} Saskatoon, SK  S7N OWO, Canada, E--mail:
{\tt szmigiel\at vincent.usask.ca}}

\leftskip=0cm   \rightskip=0cm

\vskip3\baselineskip

\centerline{\bf Abstract}

{\narrower\smallskip Different finite difference replacements for the
derivative are analyzed in the context of the Heisenberg
commutation relation.  The type of the finite difference operator
is shown to be tied to whether one can naturally consider
$D$ and $X$ to be self-adjoint and skew self-adjoint or whether they
have to be viewed as creation and annihilation operators.
The first class, generalizing the central difference
scheme, is shown to give unitary equivalent representations.
For the second case we construct a large class of examples,
generalizing previously known difference operator realizations of
$[D,X]=Id$.

\noindent PACS numbers: 3.65.--w, 3.65.Fd, 2.70.Df
\smallskip}

\vfill\eject

\noindent{\bf\Chapter. INTRODUCTION}
\vskip\baselineskip

%%%%%%%%% !!!!!!!!!!!!!!!!!!!!!!!!!!!!!!!!!!

 The idea of discrete physics and discrete space--time is a very
old one. To the best of our knowledge, the oldest reference in a
physical journal is
[\REF\AMBARCUMIAN]\footnote{$^1$}{The first questions
about the world geometry were already posed by Riemann in XIXth century
[\REF\RIEMANN].}
and it has been reconsidered many times in the past
(see [\REF\GIBBS] for an extensive bibliographical review on the
subject).
%%%%%
%The basic motivation for all those efforts was of fundamental nature:
%the quest for the origin of the fundamental length (mass) scale and
%to get rid of infinities inherently present in standard QFT
%that are especially damaging in general relativity.
 The basic motivations for all those efforts were:
(a) the presence of ultra--violet infinities in the standard
quantum field theory; (b) understanding the origin of the
fundamental length (mass) scale in the Einstein's general relativity
theory.

 In recent years there has been a growing number of attempts to find
an underlying discrete structure of space--time. At the same time
interesting discrete structures have surprisingly emerged within
originally continuous models.
The most interesting examples can be found in string theory
[\REF\KLEBANOV], Roger Penrose's
spin network calculus [\REF\PENROSEspincalc],
the loop representation of quantum gravity [\REF\LOOPREPR]
(see also [\REF\SMOLIN]),
the Bekenstein black hole entropy problem [\REF\BEKENSTEIN,
\REF\MUKHANOV, \REF\JACOBSON,
\REF\BEKENSTEINMUKHANOV, \REF\PABLOAPPb]
or the recent approach advocated by G.~t' Hooft
[\REF\THOOFTCA,\REF\THOOFTa,\REF\THOOFTb].
Also, it is interesting to mention in this context the
old work by Aharonov {\it et al} [\REF\AHARONOVPETERSEN],
where the modular variables for the coordinate and momentum
operators have been introduced to describe an infinite slit
experiment with magnetic field.

 Increasing popularity of discrete models in theoretical physics
has been stimulated by several successful attempts to discretize
continuous models for technical reasons.
This has advanced our understanding of discrete techniques.
Here, the best examples are
the lattice QCD [\REF\WILSON] and the Regge calculus
in classical gravity [\REF\REGGE].

 On the other hand, there have been several attempts to incorporate
discrete space--time at the {\it kinematical} level and to
investigate its physical consequences [\REF\SORKIN, \REF\GARAY].
This line of research can be also associated with the $q$--deformations
of the Poincare group [\REF\LUKIERPoincare].
Another interesting approach has been recently proposed
by Mazur [\REF\PABLOAPPa, \PABLOAPPb], who has given
additional physical arguments in favor of space--time discreteness.

 We assume a discrete coordinate space (see also
[\REF\PABLOPRL]) and investigate its consequences for the Heisenberg
commutation relations. This approach is purely kinematical, as the
Heisenberg relations are,  and does not depend on the details of the
underlying dynamics.
Such "unusual" realizations of the Heisenberg
algebra have also been found in 2D gravity [\REF\DOUGLAS].
These representations are not unitarily equivalent to the Schr\"odinger
representation. In particular, they cannot be exponentiated to the Weyl
form of the canonical commutation relation.

 To motivate our discussion we first consider the standard quantum mechanical
representation of the momentum
operator:
$$
p_{\mu} \ = \ -i \hbar \ {\partial\over\partial x^{\mu}}
\ . \EQNOD\MOMENTUMderivativeROLE
$$

\noindent Hence, the coordinate discretization can be implemented
through a derivative discretization:

$$
{d\over dx}
 \longrightarrow \
D_{\Delta x}
\ , \EQNOD\limit
$$

\noindent where $D_{\Delta x}$ is a discretized derivative
(difference) operator and
$\Delta x$ is the discretization parameter
(usually a fixed coordinate spacing). Even though we will call
$D_{\Delta x}$ a discretized derivative it should be noted at this
point that initially this is just some difference operator acting,
for example, on the space of smooth functions on $\Bbb R$. However,
such
operators can be naturally restricted to act on the space
of functions on $\Delta x \times \Bbb Z$.  It is in this sense that we
talk about a discrete derivative.

 We will consider a wide class of possible derivative discretization
schemes of the form:

$$
D_{\Delta x}(M+N+1) \ \equiv \ D_{\Delta x} \ \equiv \
\sum_{k=-M}^{+N} \alpha_k \ E^k_{\Delta x}
\ , \EQNOD\DERIVorig
$$

\noindent where $\alpha_k$ are real constant coefficients
and $M, N$ are integer indices corresponding to the lowest and
highest non-zero terms (hence, $\alpha_{-M} \cdot \alpha_N \ne 0$).
We take the coefficients $\alpha_k$ to be constant as one would like
to have the same definition of the discrete derivative
at all points in the coordinate space.
The expression will be called $n$--point for  $n$=$M$+$N$+$1$.
 For convenience, let us define the
shift operator $E_{\Delta x}^n$ as:

$$
E_{\Delta x}^n \ f(x) \ \equiv \ E^n \ f(x) \ = \
f(x + n \ \Delta x)
\ . \EQNOD\Eoperator
$$

\noindent One can easily check that this operator fulfills
the following commutation relations:

$$
\eqalign{
\left[ E^m , \ E^n \right] \ &= \ 0 \  , \cr
\left[ E^m , \ x \right] \ &= \ m \ \Delta x \ E^m \  , \cr
\left[ E^m , \ x \ E^n \right] \ &= \ m \ \Delta x \ E^{m+n}  \  , \cr
\left[ E^m , \ E^n \ x \right] \ &= \ m \ \Delta x \ E^{m+n}  \  . \cr
}  \EQNOD\SHIFTcomm
$$

\noindent Several authors have used a purely imaginary $\Delta x$ and
$\alpha_k$. This case is not addressed in the present paper and our
methods are not directly applicable to it.

 In addition, the formal continuum limit reads:

$$
\lim_{\Delta x \to 0} E^n_{\Delta x} \ = \ 1 \ .
\EQNOD\SHIFTlimit
$$

\noindent In the simplest case: $N = M = 1$ the difference operator
can be expressed by the Heine bracket $[\ \cdot\ ]_q$ defined for the
dimensionless variables $P, q$ as
(see Appendix~A for its basic properties):

$$
H(P,q) \ \equiv \
[\, P \, ]_q \ = \  { (q^P - q^{-P}) \over (q - q^{-1}) }
\ , \EQNOD\HEINE
$$

\noindent where $q = e^{i}$ and $P = - i {\partial\over\partial x}$.
 This notation is widely used in the context of
$q$--deformations and, in particular, to define the $q$--deformed
Poincar\'e group [\LUKIERPoincare]
(see also [\PABLOAPPa]).

%%%%%%%%%%%%%%%%%%%%%%%%%%%%%%%%%%%%%%%%%%%%%%%%%%%%%%%%%%%%%%%%%%%%%%

 This paper is organized as follows. In Sec.~2 we investigate the basic
properties of the difference operator \DERIVorig.
Sec.~3 is devoted to a construction of the Heisenberg algebra for the operator
\DERIVorig\ and its conjugate operator $X_{\Delta x}$ when the latter
is a bounded operator (finite series).
In Sec.~4 we consider the case of unbounded operators.
Sec.~5 introduces the momentum formulation of our problem and within
this representation we complete our proofs in the unbounded case in
Sec.~6.  Sec.~7 contains the summary of our results.

%%%%%%%%% !!!!!!!!!!!!!!!!!!!!!!!!!!!!!!!!!!

\vskip\baselineskip
\vskip\baselineskip
\noindent{\bf\Chapter. THE DISCRETE DERIVATIVE}
\vskip\baselineskip

 In this Section we shall investigate the general properties of the
discretized derivative \DERIVorig\ and the corresponding momentum
operator.
We would like to be able to talk about the convergence of
discretized operators to the ordinary ones when
the parameter $\Delta x$ tends to zero and
whenever we can make sense of this limit we will call it {\it the classical limit}. In particular, we stipulate that for analytic functions:

$$
D_{\Delta x} \  f(x) \ \longrightarrow \
{d\over dx} f(x)
\qquad (\Delta x \to 0)
\ , \EQNOD\limit
$$

\noindent holds pointwise in $x$. Consequently, by writing (2.1) in
terms of the translation operator
$e^{k\Delta x {d\over dx}}$ we obtain:

$$
D_{\Delta x} \  f(x) \ \equiv \
\sum_{k=-M}^{+N} \alpha_k \ e^{k\Delta x {d\over dx}} \ f(x)
\ . \EQNOD\OPNOT
$$

\noindent  Applying the Taylor expansion to \DERIVorig\ at
$x=x_{i}$ we have:

$$
\eqalign{
D_{\Delta x} \  f(x_{i}) \ = \
&f(x_{i}) \ \sum_{k=-M}^{+N} \ \alpha_k \ + \
{1\over 1!} \ {df(x_{i}) \over dx} \
\sum_{k=-M}^{+N} \ \alpha_k
\ k \ \Delta x \ +  \cr
&+ \  ...  \ + {1\over n!} \ f^{(n)}(x_{i}) \
\sum_{k=-M}^{+N} \ \alpha_k \ k^n \Delta x^n
\ + \ ...   \cr
}
\EQNOD\DEREXP
$$

\noindent From \DEREXP, we then get the
following two conditions for the coefficients $\alpha_k$
to fulfill \limit:

$$\eqalign{
\sum_{k=-M}^{+N} \ \alpha_k \ &= \ 0 \ , \cr
\sum_{k=-M}^{+N} \ k \ \alpha_k \ &= \ {1\over \Delta x} \ . \cr
} \EQNOD\CONDITIONS
$$

\noindent We may also sharpen the notion of a classical limit by
specifying the accuracy of the discretization scheme.
For example, the coefficients
$\alpha_k$  can be chosen to give the best fit to
the ordinary derivative by requiring that all
terms in \DEREXP vanish up to the order $\Delta x^{M+N}$.
As a result, in addition to \CONDITIONS, the following set of
algebraic equations for the coefficients must be satisfied:

$$
\sum_{k=-M}^{k=+N} k^n \ \alpha_k \ = \ 0 \qquad {\rm for }
\quad  n=2,3,\ldots, (M+N)
\ , \EQNOD\CONDITIONSfull
$$

\noindent and we get instead of \DEREXP\ the following approximation
for the derivative:

$$ \eqalign{
D_{\Delta x} f(x_i) \ &= \  f'(x_i) \ +
 \ {\Delta x^{M+N+1} \over (M+N+1)! }
  \  f^{(M+N+1)}(x_i) \ \times \cr
&\times \ \sum_{k=-M}^{+M} \alpha_k \ k^{M+N+1} \ + \
{\rm o}\left(\Delta x^{M+N+1}\right)
\ .  \cr  }   \EQNOD\OPTIMapprox
$$

\noindent Eqs. \CONDITIONS, \CONDITIONSfull\ comprise the set of
$M$+$N$+$1$ linear eqs. for $M$+$N$+$1$ coefficients $\alpha_k$ and they
can be solved explicitly. The corresponding determinant is the
{\it Vandermonde} determinant and its value is:
$\det \Vert A \Vert = 1! \cdot 2! \cdot \ldots \cdot (2N)! \neq 0$.
Consequently, there exists a unique solution.
Its form is given by the following:

\noindent{\bf Theorem 1:} {\it The discrete derivative of the form
\DERIVorig\ gives the best fit to the ordinary derivative,
{\it i.e.} it has expansion \DEREXP\
with the first non--vanishing coefficient for $\Delta x^{M+N+1}$,
if and only if the coefficients $\alpha_k$ are of the following
form:}

$$ \eqalign{
\alpha_0 \ &= \ {-1 \over \Delta x} \  \left[ \
\sum_{k=1}^N {1\over k} \ - \  \sum_{k=1}^M {1\over k} \
\right]  \ , \cr
\noalign{\vskip0.3\baselineskip}
\alpha_k \ &= \
{(-1)^{k+1}\over \Delta x} \  { M! \ N! \over k\ (M+k)!\ (N-k)! }
\ . }   \EQNOD\OPTIMALcoeff
$$

\noindent Higher order terms in the expansion \DEREXP\
cannot vanish, as the system would become overdetermined.

 In the symmetric case, i.e. when $M=N$, Theorem 1 implies
the symmetry between forward and backward terms.  Indeed, in this case
the coefficients \OPTIMALcoeff\ have the following simple
properties:

$$
\alpha_k(\Delta x) = - \, \alpha_{-k}(\Delta x)=\alpha_{-k}(-\Delta x)
\ , \EQNOD\antisymm
$$
Moreover, for a fixed $k$,
$$
\lim_{N\to\infty} \alpha_k (\Delta x)\ = \ (-1)^{k+1} \ \, {1\over (\Delta x)k}
\qquad ( k \neq 0 )
\ . \EQNOD\coefflimit
$$
\noindent The meaning of (2.9) is explained in Appendix~B.
\noindent Moreover, eq. \antisymm\ implies the symmetry of
the discrete derivative:

$$
D_{\Delta x} \ = \ + D_{-\Delta x}
\ . \EQNOD\DERIVsymmetry
$$

\noindent The operator $D_{\Delta x}$ satisfying \DERIVsymmetry\ we
shall call {\bf symmetric} (in $\Delta x$).
 Later on we shall limit ourselves to this case as it preserves the
parity symmetry.
Discretization schemes and corresponding coefficients satisfying,
in addition to \CONDITIONS,
\CONDITIONSfull\ and symmetric we shall call
{\bf optimal}.
In fact, in the special symmetric case $N=1$ \antisymm\
implies that the derivative is optimal and we obtain the well known
central difference scheme.
The numerical values of the optimal
$\alpha$--coefficients for the lowest  $(2N+1)$--point
schemes are given in the {\it Table 1} below.

\midinsert
\vskip\baselineskip
\noindent {\it Table 1. Values of the lowest optimal
$\alpha$--coefficients.}
$$
\vbox{
\settabs
\+ x & xxxxxxxxx & xxxxxxx & xxxxxxx & xxxxxxx & xxxxxxx & xxxxxxx &
xxxxxxx &\cr
\hrule height1pt
\vskip4pt
\+ & $N=M$\hfill & $\alpha_1$ & $\alpha_2$ & $\alpha_3$ & $\alpha_ 4$ &
$\alpha_5$ & $\alpha_6$ &\cr
\vskip4pt
\hrule
\vskip4pt
\+ & \hfil 1 \hfil &  1/2  &        &       &        &       &      &\cr
\+ & \hfil 2 \hfil &  2/3  & --1/12 &       &        &       &      &\cr
\+ & \hfil 3 \hfil &  3/4  & --3/20 & 1/60  &        &       &      &\cr
\+ & \hfil 4 \hfil &  4/5  & --1/5  & 4/105 & --1/280 &      &      &\cr
\+ & \hfil 5 \hfil &  5/6  & --5/21 & 5/84  & --5/504 & 1/1260 &    &\cr
\+ & \hfil 6 \hfil &  6/7 & --15/56 & 5/63  & --1/56 & 1/385 & --1/5544 &\cr
\vskip4pt
\hrule height1pt
 }   $$
\vskip0.5\baselineskip
\endinsert

 In analogy to \OPNOT, the optimal derivative can be rewritten as:

$$
D_{\Delta x} \  f(x) \ \equiv \ 2 \ \sum_{k=1}^{N}
\alpha_k \ \sinh\left(k\Delta x {d\over dx}\right)  \ f(x)
\ . \EQNOD\OPTIMOP
$$

\noindent In particular, for the optimal 3--point (= central difference)
discretization sche\-me we have:

$$
D_{\Delta x} \  f(x) \ = \
{1 \over \Delta x} \ \sinh\left(\Delta x {d\over dx}\right)  \ f(x)
\ = \
{1\over i\ \Delta x} \ \sin\left(i \ \Delta x {d\over dx}\right)
\ f(x) \ .
\EQNOD\sinhDER
$$

\noindent The representation \sinhDER\ is
exactly the derivative represented by the Heine symbol
\HEINE\   ([\LUKIERPoincare, \PABLOAPPa]).

%%%!!!!!
 Originally, Heisenberg introduced his matrix mechanics
by postulating that mathematical operations in classical equations of
motion should be reinterpreted (that has led to the correspondence
principle). In this spirit Dirac has postulated that the "quantum
differentiation" must satisfy the additivity and Leibniz rule. From this
he has obtained the fundamental (Heisenberg) commutation relations for
coordinates and momenta [\REF\DIRACQUANT].
We would like to proceed in an analogous way. However, in our case
%%%%%%%%
not all general rules for derivatives are satisfied by \DERIVorig,
subject to \CONDITIONS.
The basic properties of the ordinary derivative are
listed below along with some comments on how those properties change if
one uses $ D_{\Delta x} $ instead of $d\over dx$:

\item{1.} Additivity. Ordinary and discrete derivative are both
linear :

$$D_{\Delta x} \  [ f(x) + g(x) ] =
D_{\Delta x} \  f(x) + D_{\Delta x} \  g(x) \ . $$

\item{2.} Leibniz rule. The rule for differentiation of the
product is not satisfied exactly by \DERIVorig.
The deviation is of order ${\rm O}\left(\Delta x\right)$
and for the optimal scheme of order
${\rm O}\left(\Delta x^{2N}\right)$.

\item{3.} Derivative of the composite function. The same
result as for the Leibniz rule holds for the derivative of a
composite function:
$D_{\Delta x} [ f\circ g(x) ] = D_{\Delta x} f(g(x)) \ \times \
D_{\Delta x} g(x) \ + \ {\rm O}\left(\Delta x^{2N}\right)$.

\item{4.} Derivative of monomials. Discrete derivative of the
monomial $x^n$ is identical to the ordinary derivative
if the discretization scheme is optimal and the exponent
$n \le 2N$. In general, assuming \CONDITIONS\ and \antisymm\
we have:

$$
D_{\Delta x}(N) \ x^n \ = \ n \ x^{n-1} \ + \
\sum_k \sum_{i=2}^n \alpha_k \ k^i \ \left( {n\atop k} \right) \
\Delta x^i \ x^{n-i}
\ . \EQNOD\monomialDER
$$

\item{5.} {Heisenberg commutation relation. This will be the subject
of the following sections.
We shall
present a construction of the discrete counterpart of the Schr{\"o}dinger
couple $x$ and ${d\over dx}$
satisfying a discrete
analog of the continuous Heisenberg commutation relation:
$[{d\over dx}, x ] f(x) = f(x)$.
Naively, that is, by replacing only ${d\over dx}$ with $D_{\Delta x}$
 and using \DERIVorig, we get:

$$ [ D_{\Delta x}, \ x ] \ f(x) \ = \ \Delta x \ \sum_k
k \ \alpha_k \ f(x+k \Delta x )
\ . \EQNOD\DtCOMMUTATOR
$$
For the optimal discretization we will construct a position operator called
$X$ which together with $D_{\Delta x}$ forms a conjugate pair, that is the
relation $ [ D_{\Delta x},X ]=I$ holds on a dense domain in a Hilbert space}.

\item{6.} Hermicity. In the symmetric case we have a natural Hermitian conjugation induced from the relation $(E^n_{\Delta x})^\dagger=E^{-n}_{\Delta x}$ The operator $D_{\Delta x}$, like the operator
$d\over dx$ is formally anti-Hermitian if $M=N$ and $\alpha_k(\Delta x)
=-\alpha _{-k}(\Delta x)$ and we have:

$$
\int \left( D_{\Delta x} f(x) \right) \ g(x) \ dx \ = \
- \ \int f(x) \ \left( D_{\Delta x} g(x) \right) \ dx
\ . \EQNOD\ANTIHERMICITY
$$

\noindent A more detailed analysis will be given in the subsequent
sections.

\vskip\baselineskip
\vskip\baselineskip
\noindent{\bf\Chapter. LIE ALGEBRAIC DISCRETIZATION}
\vskip\baselineskip

 The notion of the {\it Lie algebraic discretization} has been introduced
in [\REF\TURBINERSMIR].  It can be described as a special case of
representation theory of the enveloping algebra of the Heisenberg algebra
in which at least one of the generators is a difference operator.
As a result, some linear difference equations have a representation
theoretic meaning similar to the D-module interpretation of linear
differential equations [\REF\COUTINHO],
in particular in Sec.~6. The most important application
seems to be
to describe quasi-polynomial solutions
 to a certain class of difference equations.  On the other hand the
whole
procedure has certain appeal of {\it d\'{e}j\`{a} vue} in the context
of basic quantum mechanics and from this point of view one can regard
the authors' concept of the Lie algebraic discretization as a
quantization procedure.

 Following [\TURBINERSMIR], we start with
\MOMENTUMderivativeROLE\
and we look for a conjugate  operator $X\equiv X_{\Delta x}$
that satisfies:

$$
\left[ \ D_{\Delta x} , \ X \ \right] \ = \ 1
\ , \EQNOD\HEISENBERGrelation
$$

\noindent with the additional condition (`classical limit'):

$$
\lim_{\Delta x \to 0} \ X_{\Delta x} \ = \ x
\ . \EQNOD\xOPERATORlimit
$$

\noindent In [\TURBINERSMIR], only the special cases of
the forward and backward {\it Euler schemes}
that are not symmetric under reflections \DERIVsymmetry\
have been discussed.  This can be improved in two ways.  Firstly,
we can start with the more general discretization scheme \DERIVorig.
Secondly, we can require that $D_{\Delta x}$ be formally skew-Hermitian
whereas  $X_{\Delta x}$ be formally Hermitian relative to some $\star$
anti-involution to ensure a natural physical interpretation.

In [\TURBINERSMIR], it has been shown that for derivatives in the form:

$$
D_{\Delta x}^+ f(x) \equiv {f(x+\Delta x) - f(x) \over \Delta x} \ ,
 \qquad
D_{\Delta x}^- f(x) \equiv {f(x) - f(x-\Delta x) \over \Delta x} \ ,
\EQNOD\TurbDER
$$

\noindent we get the Heisenberg algebra for $D^+$ and
$x (1-\Delta x D^-)$:

$$
[ D^+, \ x (1-\Delta x D^-) ] = 1
\ . \EQNOD\TurbHEISENBERG
$$

 However, as has been mentioned in Sec.~2,
we are not satisfied with the Euler forms of the
discrete derivative \TurbDER, as these schemes
are not optimal and symmetric
with respect to reflections\footnote{$^2$}{It can be
shown that for classical dynamical systems the optimal schemes
are closer to continuous derivative with respect to the
stability analysis and they preserve the Hamiltonian structure
of the phase space, as well [A.Z.~G\'orski and J.~Szmigielski,
in preparation].}.
In particular, this form of the discrete derivative does not seem to
be skew-Hermitian with respect to any natural Hermitian structure.
%%%!!!!!
Also, the central difference scheme (and optimal, in general)
was shown to enable a consistent formulation of the action principle and
the Euler--Lagrange equations in the classical theory [\REF\COLE].
%%%%%%%%
Another possibility would be that the pair
$D_{\Delta x}, X$ is a pair of conjugate operators
(annihilation and creation operators).  But here again, no natural
Hermitian structure on the Hilbert space of states seems to induce that.

 First, we take up a question of generalizing the results of
[\TURBINERSMIR] to a wider class of discrete
operators. As the starting point we generalize the form of the
coordinate operator used in [\TURBINERSMIR], stipulating the following
{\it Ansatz} :

$$
X \ \equiv \ \sum_{k=-N}^M \beta_k \ x\ E^k \ ,
\EQNOD\Xdef
$$

\noindent subject to the condition (continuous limit):

$$
\sum_{k=-N}^M \beta_k \ = \ 1 \ .
\EQNOD\Xlimit
$$

\noindent In fact, one can add to \Xdef\ two additional terms without
changing the conclusions.  Thus $X$ might be taken to be of the form:

$$
X \ \equiv \ \sum_{k=-N}^M \beta_k \ x\ E^k \  +
\ \sum_{k=-N}^M \beta_k' \ E^k \ x \  +
\ \sum_{k=-N}^M \beta_k'' \ E^k \  .
\EQNOD\XGENdef
$$

\noindent The last term commutes with the derivative $D_{\Delta x}$
and does not change commutation relation \HEISENBERGrelation.
It changes however condition \Xlimit\ replacing it with:

$$
\sum_{k=-N}^M \left(\beta_k + \beta_k' \right) = 1 \ ,
\qquad \sum_{k=-N}^M \beta_k'' = 0 \ .
\EQNOD\XlimitGEN
$$

 The second term in the right hand side of \XGENdef, due to
\SHIFTcomm, has the same commutation relation as \Xdef\
leading to the substitution:

$$
\beta_k \ \longrightarrow \  \beta_k + \beta_k'
\ . \EQNOD\betaSUBST
$$
From now on, we restrict our attention to \Xdef.
From \HEISENBERGrelation\ we get the following conditions for the
$\beta$`s:

$$\eqalign{
\sum_{k=-N}^M \beta_k \ &= \ 1 \ , \cr
\Delta x \ \sum_{k=-M\atop -N \le n-k \le M}^N
\alpha_k \ \beta_{n-k} \ k \ &= \ \delta_{n0}
\ , \cr
} \EQNOD\betaEQS
$$

\noindent where $n$ goes from $-(M+N)$ to $+(M+N)$. Here,
\betaEQS\ is a set of $(2M+2N+2)$ eqs. for
$(2M+2N+2)$ coefficients $\alpha_k, \beta_k$ plus
two additional equations \CONDITIONS.
Hence, the system is formally overdetermined.
Let us start with the second equation above (set of $2N+2M+1$ eqs.)
and rewrite it as two separate sets, each of size $M+N+1$,
corresponding to $n\le 0$ and $0\le n$ respectively.
For simplicity we introduce matrix notation in the
$(M+N+1)$--dimensional space of coefficients.
Let us define vectors:

$$
\beta \ = \
\left[ \matrix{ \beta_{-N} \cr  \beta_{-N+1} \cr \vdots \cr
\beta_{M} \cr } \right] \ , \qquad
b \ = \
\left[ \matrix{ 1/\Delta x \cr  0  \cr \vdots \cr  0 \cr } \right]
\ , \EQNOD\bVECTS
$$

\noindent and two $(N+M+1)\times(N+M+1)$ matrices:

$$ \eqalign{
A_- \ &= \ \left[ \matrix{
N\alpha_N & \ldots &  \ldots & -M\alpha_{-M} \cr
(N-1)\alpha_{N-1} & \ldots  & -M\alpha_{-M} & 0 \cr
\ldots & \ldots  & \ldots & \ldots \cr
-M\alpha_{-M} & 0  & \ldots & 0 \cr     }
\right] \ ,   \cr
\noalign{\vskip0.3\baselineskip}
A_+ \ &= \ \left[ \matrix{%
N\alpha_N & \ldots &  \ldots & -M\alpha_{-M} \cr
0 & N\alpha_N & \ldots & (-M+1)\alpha_{-M+1} \cr
\ldots & \ldots & \ldots & \ldots \cr
0 & \ldots & 0 & N\alpha_N \cr     }
\right] \ .   \cr
}   \EQNOD\Amatrices
$$

\noindent Now, our equations can be rewritten as two sets of
equations for the coefficients $\beta_k$ in the following simple form:

$$
A_- \ \beta \ = \ b \ , \qquad
A_+ \ \beta \ = \ b \ ,
\EQNOD\matrixEQS
$$

\noindent where the first equations in both sets are identical.
The determinants of the systems \matrixEQS\ are:

$$
\hbox{\rm det} A_- \ = \ (-M \ \alpha_{-M})^{M+N+1}
\ , \qquad
\hbox{\rm det} A_+ \ = \ (N^{M+N+1} \ \alpha_N)^{M+N+1} \ .
\EQNOD\Adets
$$

\noindent and \matrixEQS\ can be solved explicitly.
Indeed, there are three possibilities to consider:
1. both determinants are nonzero ({\it i.e.} $M\cdot N \ne 0$),
2. one of the determinants is zero, then either $M=0$ or $N=0$,
3. both determinants are zero ($M=N=0$).
In the last case the derivative expansion \DERIVorig\
consist of one nonzero term ($\alpha_0 E_0$) only and
the second condition of \CONDITIONS\ cannot be fulfilled.
In the first case, using Cramer`s rule,
we get two different solutions to \matrixEQS, hence the system
is inconsistent.
The only nontrivial possibility is to have the second case.
Let us assume that $\alpha_N\ne0$. Then from the second
equation of \matrixEQS\ we get the following (unique) solution
for $\beta$'s:

$$
\beta_{-N} = {1\over N\alpha_N \Delta x} \ ,
\quad \beta_{-N+1} = \ \ldots \ = \beta_M = 0 \ ,
\EQNOD\betaSOL
$$

\noindent and the first equation of the set \betaEQS\ implies:
$\beta_{-N} = 1$ and $\alpha_N = 1/N\Delta x$.
Now, consistency of the first eq. of \matrixEQS\ implies
that the only non--zero $\alpha$'s can be:
$\alpha_0$ and $\alpha_N$.
Finally, eqs. \CONDITIONS\ are satisfied as well if:
$\alpha_0 = - \alpha_N$.
Identical solution can be found for $\alpha_{-M}\ne 0$.
Hence, we have got the following

\noindent{\bf Theorem 2:}
{\it For the discretized derivative operator \DERIVorig\
satisfying \CONDITIONS\ and the coordinate operator in the form
\Xdef\ with condition \xOPERATORlimit\ to satisfy Heisenberg
commutation relation \HEISENBERGrelation\
the coefficients $\alpha_k, \beta_k$ must be either in
the forward Euler form:

\vskip-0.3\baselineskip
$$
\alpha_k \equiv \alpha^F_k = -{1\over N\,\Delta x} \delta_{k0} +
{1\over N\,\Delta x} \delta_{N,k} \ ,
\qquad
\beta_k \equiv \beta^F_k =  +\delta_{-N,k}
\ , \EQNOD\HEISconF
$$

\vskip-0.3\baselineskip
\noindent or in the backward Euler form:
\vskip-0.3\baselineskip

$$
\alpha_k \equiv \alpha^B_k = {1\over M\,\Delta x} \delta_{k0} -
{1\over M\,\Delta x} \delta_{-M,k} \ ,
\qquad
\beta_k \equiv \beta^B_k = +\delta_{-M,k}
\ . \EQNOD\HEISconB
$$
}

Thus, for $X$ of the form \Xdef\ the only way to satisfy
\HEISENBERGrelation\
is to take $D_{\Delta x}$ as the Euler derivatives \TurbDER.
In this case the coordinate
operators are:

$$
X^{B,F} \ = \ x \ (1 - \Delta x \ D_{N\Delta x}^{-\atop+} ) \ = \
x \ E_{\Delta x}^{\pm N}    \ .
\EQNOD\TurbXoper
$$

\noindent This proves that the Turbiner--Smirnov discretization scheme
is unique under conditions \HEISENBERGrelation, \xOPERATORlimit,
\Xdef.
Hence, an optimal derivative with the discretized coordinate
operator of the form \Xdef\  cannot satisfy the Heisenberg
commutation relation.
In fact, this is impossible for any symmetric
discretization scheme with finite series in \Xdef.

\vskip\baselineskip
\vskip\baselineskip
\noindent{\bf\Chapter. TRANSCENDENTAL OPERATORS $X$}
\vskip\baselineskip

 In the previous Section we have limited our considerations
to the coordinate operators \Xdef\ in the form of a finite
series. However, there is still a possibility of having the
Heisenberg algebra with the discrete derivative
of the form \DERIVorig\ and an infinite series for $X$.
We shall investigate the algebraic aspects of this problem in
this section and a complete, rigorous solution of the problem
will be given in Sec.~6.
We should mention that the operator $X$
is always unbounded and eventually a special care needs to be
exercised when dealing with any expressions involving $X$. In this
section we study only formal aspects of our problem leaving aside
important questions like the domain of the definition of $X$ or
in what sense we understand expressions below involving infinite
sums.  We will come back to those questions in Sec.~5 and 6.

In the present section we postulate that our coordinate operator
has the following form:

$$
X \ =  \ x \ \sum_{k=-\infty}^\infty   \beta_m \ E^m_{\Delta x}
\ , \EQNOD\XdefINF
$$

\noindent and we have an infinite set of linear algebraic
eqs. \betaEQS\ for $n=0, \pm1, \pm2, \pm3, ...$ .
In addition, the continuous limit condition \Xlimit\ is
tentatively taken to mean:
$$
\sum_{k=-\infty}^{+\infty} \ \beta_k \ = \ 1 \ .
\EQNOD\XlimitINF
$$
Eventually, we will have to regularize this condition since in many
examples a direct
evaluation of this limit does not make sense.

 Defining infinite dimensional matrices
and vectors analogous to \bVECTS, \Amatrices:

$$
\beta_+ \ = \
\left[ \matrix{ \beta_{-N} \cr  \beta_{-N+1} \cr \beta_{-N+2} \cr
\vdots \cr
} \right] \ , \qquad
\beta_- \ = \
\left[ \matrix{ \beta_{M} \cr  \beta_{M-1} \cr \beta_{M-2} \cr
\vdots \cr
} \right] \ , \qquad
b \ = \
\left[ \matrix{ 1/\Delta x \cr  0  \cr 0 \cr \vdots \cr} \right]
\ , \EQNOD\bVECTSinf
$$

$$ \eqalign{
A_+ \ &= \ \left[ \matrix{
N\alpha_N & \ldots &  -M\alpha_{-M}
& 0 & \ldots & \ldots \cr
0 & N\alpha_{N} & \ldots  & -M\alpha_{-M}
& 0 & \ldots \cr
\ldots & \ldots  & \ldots & \ldots & \ldots & \ldots \cr }
\right] \ ,   \cr
\noalign{\vskip0.3\baselineskip}
A_- \ &= \ \left[ \matrix{
-M\alpha_{-M} & \ldots &  N\alpha_{N} & 0 & \ldots & \ldots \cr
0 & -M\alpha_{-M} & \ldots  & N\alpha_{N} & 0 & \ldots \cr
\ldots & \ldots & \ldots & \ldots & \ldots & \ldots \cr }
\right] \ ,   \cr
}   \EQNOD\AmatricesINF
$$

\noindent formally, we get, instead of \matrixEQS\ the following two
sets of eqs.:

$$
A_+ \ \beta_+ \ = \ b \ , \qquad
A_- \ \beta_- \ = \ b \ .
\EQNOD\matrixEQSinf
$$

 In both sets the first equation is identical and relates at most
$(M+N+1)$ coefficients $\beta_m$.
Each next equation involves one more (higher order) coefficient
that can be explicitly computed.
Hence, our system has $(M+N)$ free parameters.
In the special case, when $M\cdot N=0$ the first or last
element in each row in \AmatricesINF\ vanishes and we have
$(M+N-1)$ free parameters. This implies for the kernel of
operators $A_+, A_-$:

$$
D \ \equiv \ \hbox{\rm dim ker } A_+ \ = \
\hbox{\rm dim ker } A_- \ = \ M + N - \delta_{MN, 0}
\ . \EQNOD\DimKerA
$$

\noindent Hence, we have proven the following

\noindent{\bf Theorem 3:}
{\it For operator $X$ in the form of infinite series
\XdefINF\ there is $(M+N-\delta_{MN,0})$--parameter
family of formal solutions to the Heisenberg relation
\HEISENBERGrelation.}

\smallskip
\noindent
{\bf Remark:}  We do not claim at this point that any of the formal
solutions satisfy \XlimitINF.
\smallskip
In the case $M\cdot N \ne 0$ eq. \matrixEQSinf\
imply one algebraic condition for the coefficients
$\{\beta_{-N}, \ \ldots, \beta_M \}$:

$$
\sum_{k=-M}^{+N} k \ \alpha_k \ \beta_{-k} \ = \ {1\over \Delta x}
\ , \EQNOD\genBETAeq
$$

\noindent and the higher coefficients $\beta_k$ ($k>M$ and $k<-N$)
can be uniquely determined from:

$$\eqalign{
\beta_{M+i} \ &= \ {-1\over M\alpha_{-M} } \
\sum_{k=-M+1}^{+N} k \ \alpha_k \ \beta_{i-k} \ , \cr
\beta_{-(N+i)} \ &= \ {-1\over N\alpha_{N} } \
\sum_{k=-M}^{N-1} k \ \alpha_k \ \beta_{-(i+k)} \ , \cr
} \EQNOD\genBETAsol
$$

\noindent where $i = 1, 2, \ldots$ with additional
condition \XlimitINF.

 In the remainder of this section we shall assume that the coefficients
$\alpha_k$ are optimal and the coefficients $\beta_k$ have
the symmetry:

$$
\beta_k \ = \ \beta_{-k} \ , -N\le k \le N \, . \EQNOD\BETAsymmetry
$$

\noindent This implies the symmetry of the discrete coordinate
operator with respect to the $\Delta x$ reflections:

$$
X_{\Delta x} \ = \ X_{-\Delta x} \ ,
\EQNOD\Xsymmetry
$$
in analogy to the discrete derivative operator
\DERIVsymmetry.

 Indeed, now we have $N+1$ $\beta$--parameters
$\{\beta_0, \beta_1, \beta_2, \ldots, \beta_N \}$
subject to :

$$
\sum_{k=1}^{N} k \ \alpha_k \ \beta_k \ = {1\over 2 \Delta x}
\ . \EQNOD\genBETAeqSYMM
$$

\noindent The higher order coefficients
($\beta_k$ for $\vert k\vert > N$)
are uniquely determined from:

$$
\beta_{N+i} \ = \ {-1\over N\alpha_N} \ \sum_{k=1}^N
k \ \alpha_k \ \beta_{i-k} \ - \ {1\over N\alpha_N} \
\sum_{k=1}^{N-1} k \ \alpha_k \ \beta_{i+k}
 \  , i = 1, 2, \ldots \ .  \EQNOD\genBETAsolSYMMETRIC
$$

\noindent Using induction on $i$ and \genBETAeq\ we get the claim:

$$
\beta_{-(N+i)} \ = \ \beta_{N+i}
\ . \EQNOD\genBETAsolSYMMETRICnegative
$$

\vskip0.5\baselineskip

We remark that the normalization condition
\XlimitINF\ can be rewritten as:

$$
2\ \sum_{k=1}^{+\infty} \ \beta_k \ = \
1 - \beta_0 \ ,
\EQNOD\XlimitINFsymmetric
$$

\noindent{\it i.e.} we are left with $(N-1)$ free parameters $\beta$.
All considerations in this section were rather formal as we
deal with infinite series whose convergence is unclear.
In fact, as we shall see below the series might diverge in the usual
sense.
We discuss in details a concrete example to illustrate this point.

 In the case $N=1$ \genBETAsolSYMMETRIC\ implies
the following condition for the
$\beta$--coefficients:

$$
\beta_{i+2} \ = \ \beta_i \ , \qquad i=0, 1, 2, \ldots \ .
\EQNOD\betaNone
$$

\noindent Hence, we have two free parameters: $\beta_0$ and $\beta_1$.
The latter can be determined from \genBETAeqSYMM, giving:

$$
\beta_1 \ = \ + 1 \ .
\EQNOD\betaONE
$$

\noindent This provides an exact formula for all odd coefficients
in the following compact form:

$$
\eqalign{
\beta_{2k+1} \ &= \ ( -1 )^k \ , \qquad k=0, 1, 2, \ldots \ , \cr
\beta_{-(2k+1)} \ &= \  \beta_{2k+1}  \ ,  \cr
}  \EQNOD\betaODD
$$

\noindent and for even coefficients we have:

$$
\beta_{2k} \ = \ (-1)^k \ \beta_0 \ .
\EQNOD\betaEVEN
$$

\noindent It is clear from \betaODD\ that our series is indeed
divergent and we must redefine the classical limit to make sense
of the summation
in \XdefINF.
There is no unique way of doing that.  We will say more about this in the next  section.

 At the moment, we are still left with one free parameter,
$\beta_0$, and one equation, the normalization condition
\XlimitINFsymmetric.
As a special case we can put

$$
\beta_0 \ = \ 0 \ .
\EQNOD\betaZERO
$$

\noindent Hence, a formal solution for the
discretized coordinate operator, $X$, in the $N=1$ case is
\XdefINF\ with $\beta$--coefficients given by:

$$
\eqalign{
\beta_{2k} \ &= \ 0 \ , \qquad \forall k \cr
\beta_{2k+1} \ &= \ (-1)^k \ , \qquad k=0,1,2,\ldots \ , \cr
\beta_{-(2k+1)} \ &= \ \beta_{2k+1} \ , \qquad k=0,1,2,\ldots \ . \cr
}
\EQNOD\betaNoneFINAL
$$

\vskip\baselineskip

 Now, this solution can be compared with the solution
found by Frappat and Sciarino [\REF\FRAPPAT].
Their solution, after restricting it to 1 dimension, reads:

$$
X \ = x\ {1\over E + E^{-1} } +{1\over E + E^{-1}} x \ .
\EQNOD\FRAPPATsolution
$$

\noindent There is a formal analogy between their solution and
\betaNoneFINAL. To see that analogy one expands the first term in
\FRAPPATsolution\ at $E=0$ and the second one at $E^{-1}=0$. There is, however, an important
difference between the approach of the present paper and that of
[\FRAPPAT].
In their approach, the discretization parameter $\Delta x$
is purely imaginary.  As a consequence, the shift operator $E$ is formally
Hermitian and positive definite, while the momentum operator corresponding to
our $D_{\Delta x}$ is unbounded. Moreover, since the shift is in the
imaginary direction it is hard to see what this has to do with a discrete
space-time.  By contrast, we are treating the shift
as a unitary operator, thus
retaining the real discretization parameter, which, in turn, ensures that
the momentum is bounded.

\vskip\baselineskip
\vskip\baselineskip
\noindent{\bf\Chapter. MOMENTUM FORMULATION}
\vskip\baselineskip

 In this section we reformulate some of the results obtained earlier
as well as clarify some of the points left obscure in our discussion of
equation \FRAPPATsolution.

The basic idea is to realize the algebra of shift operators
as an algebra of multiplication operators on the Hilbert
space of square integrable functions on the circle.  One can also view
this realization as the momentum space counterpart of the content of
the preceding sections.  Throughout this section $\Delta x=1$.

At the core of the idea of discretization is that in all operations
one proceeds by performing a sequence of discrete steps of size,
in this section, 1.  This calls for bringing the Fourier transform
into the picture.  To this end we will first reinterpret
\SHIFTcomm.  Our setup is as follows.  We consider the unit circle:

$$
S^1=\{z \in \C: \ |z|=1\} \EQNOD\circleDEF
$$

\noindent and the Hilbert space $\H=\L^2(S^1)$.  We recall

$$
L^2(S^1) = \left\{ f=\sum _{-\infty}^{\infty}a_n z^n:\,
\sum _{-\infty}^{\infty}|a_n|^2<\infty \right\} \ .
\EQNOD\HilbertSpaceDEF
$$

 We use the following parametrization of the circle $S^1$: $z=\exp i
\theta, \quad \theta \in \R \pmod {2 \pi}$.
Thus a function $f \in L^2(S^1)$ can be written as:

$$
f(e ^{i\theta}) \ = \ \sum _{n \in \Z}\, a_n \ e ^{in \theta} \ ,
\EQNOD\LtwoFUNCTION
$$

\noindent and $a_n$ are the Fourier coefficients of $f$.
Since the shift operator $E$ defined in \Eoperator\ corresponds to the
elementary step of unit size we
postulate that $E=T_z$, where $T_z$ is the operator of multiplication by
$z$. We will say more about multiplication operators towards
the end of this section.  Till further notice we also suppress writing $T$.
Thus we set

$$
E^n=z^n \ , \quad  x = -z \, {d \over dz} \ , \quad
n \in \Z  \ .
\EQNOD\SHIFTreformulate
$$

\noindent It is routine to check that \SHIFTcomm\ hold.  Furthermore we
define

$$
D(z) \ :=\sum _{k=-M}^N \alpha _k \ z^k \ .
\EQNOD\DOPdef
$$

In other words the difference operator becomes now a multiplication by
a Laurent polynomial in $z$. We might observe that the operator $x$
above is one of the generators of the Virasoro algebra (with the trivial
central term), called, $L_0$.
Now we reinterpret equations \CONDITIONS\ and \CONDITIONSfull.
We see that \CONDITIONS\ goes into

$$
D(z=1)=0 \ , \qquad (L_0 D)(z=1)=-1 \ .
\EQNOD\CONDITIONSreformul
$$

\noindent Similarly, \CONDITIONSfull\ goes into

$$
(L_0^n D)(z=1)=0 \ , \qquad \hbox{{\rm for}} \ \ n=2,3,\dots , (M+N)
\ .   \EQNOD\CONDITIONSfullREFORMUL
$$

\noindent Now, it is immediate that the coordinate operator $X$ should
be sought (see \TurbHEISENBERG) in the form

$$
X=-z\sum_{n=-\infty}^{\infty}\beta _n \ {d \over dz}z^n \ ,
\EQNOD\reformul
$$

\noindent where

$$
\sum_{n=-\infty}^{\infty}\beta _n =1 \ .
\EQNOD\betaCONDreformul
$$

\noindent Let us now, at least formally, introduce the function
$h(z)=\sum_{-\infty}^{\infty}\beta _n z^n$.  From the above condition
we get that $h(1)=1$.
Moreover $X$ can be written

$$
X(z)=L_0 h(z)=-zh(z){d \over dz}-zh^{\prime}(z) \ .
\EQNOD\XzONE
$$

\noindent As was already explained earlier, we can in fact ignore the
second part and look for $X$ in the form:

$$
X(z)=-f(z) \ {d \over dz} \ , \qquad f(1)=1 \ .
\EQNOD\XzTWO
$$

\noindent Thus $X$ is formally an element of the Virasoro algebra.
We want that

$$
\left [D(z),X(z)\right ]=\left [f(z){d \over dz},D(z)\right ]=1 \ ,
\EQNOD\DzCOMMUTATOR
$$

\noindent where $D(z)$ is a function (Laurent polynomial) defined
previously. We therefore obtain that:

$$
f(z) \ D^{\prime}(z)=1 \ ,
\EQNOD\fzEQ
$$

\noindent which formally implies

$$
f(z)={{1}\over{D^{\prime}(z)}} \ .
\EQNOD\fzEQsolv
$$

\noindent We note that the condition $f(1)=1$ is automatically satisfied
as a consequence of the second condition imposed on $D$.
However, it is important to underscore the fact that the latter
expression can only be understood formally at this stage of
our analysis. Yet, there are some special cases when we do not expect
to have any difficulties.  For example, if $D(z)=A+Bz$, that is
$M=0,\, N=1$.  Then we impose the following conditions:

$$
D(1)=0\ , \quad   D^{\prime}(1)=1 \ ,
\EQNOD\DboundCOND
$$

\noindent which gives $D(z)=z-1$ and, consequently, $f(z)=1$.  Thus we get the pair $D(z)=z-1,\, X(z)=-{{d} \over {dz}}$ which
clearly satisfies \DzCOMMUTATOR.  Using \XzONE\ we get that in this case $h(z)=
{{1}\over {z}}$.  Hence $\beta _{-1}=1$ and $\beta _i=0, i\ne -1$.
Going back to the difference operator realization we get
that $D=E-I, \, X=x E^{-1}$ which is
the pair found in [\TURBINERSMIR] for $\Delta x=1$.
We can contrast this example with that of the symmetric discretization
scheme for which $D(z)={{1}\over {2}}(z-z^{-1})$. In that case
$f(z)={{2 z^2}\over {1+z^2}}$
which is not bounded on the unit circle and consequently $f(z)$ is
not a multiplier (that is the multiplication by $f$ is not a bounded
operator on $L^2(S^1))$. This example is explained in detail in the
next section.

However, there are many cases for which
$f$ is a multiplication operator on the circle.
We present now a general
construction of $X$ for which $f$ is a multiplier. One can view this
construction as a generalization of [\TURBINERSMIR].

First, we review a few elementary facts from the theory of doubly
infinite Toeplitz operators.  Initially, we can consider a vector space
of sequences
$a=\{a_n\}_{n\in \Z}$ such that
$\sum _{n\in \Z} |a_n|^2 < \infty$.
This space, called $\ell ^2$, is isomorphic to $L^2(S^1)$,
the fact well known from the theory of Fourier series.
On $\ell ^2$ we introduce a linear operator $T_c$:

$$
a_m \rightarrow  \sum _{n\in {\Bbb Z}}c_{m-n}a_n \ ,
\EQNOD\ToeplitzOPERATORdouble
$$

\noindent where $c=\{c_n \}_{n\in \Z}$ is a sequence of numbers.
It is known, that $T_c$ is a bounded operator on $\ell ^2$ iff $c$
is a sequence of
the Fourier coefficients of an essentially bounded function on $S^1$.
We recall that $\phi (\theta)$ is an essentially bounded function on
$S^1$ if it is bounded almost everywhere there.  The isomorphism between
$\ell ^2$ and $L^2$ allows one to study the Toeplitz operator $T_c$ as
acting on $L^2(S^1)$.
We will say that $\phi(\theta)$ is a symbol of $T_c$ if
under this isomorphism $T_c$ acts on $L^2(S^1)$ as:

$$
(T_c f )(\theta)=\phi (\theta) f(\theta) \ ,
\qquad f\in L^2(S^1) \ .
\EQNOD\ToeplitzACTING
$$

\noindent Because of this relation we often write $T_{\phi}$ to
denote the Toeplitz operator with symbol $\phi$.
We will need in this section one more result from the theory of
Toeplitz operators.  The result below gives a complete
description, in terms of symbols, of invertible Toeplitz operators.
\smallskip
\noindent{\bf Lemma 1:}\quad {\it $T_{\phi}$ is invertible if and only
if $1/{\phi}$ is essentially bounded. If this holds,
$T^{-1}_{\phi}=T_{1/{\phi}}$.}

Now we apply this lemma to our problem.  Let us first consider
a Laurent polynomial $g(z)$ such that $Res _{z=0} g =0$ and $g$ is
nowhere zero on $S^1$.  We can normalize such a $g$ to satisfy $g(1)=1$.
Now we define

$$
D(e ^{i\theta}):=i\int _0 ^{\theta}e ^{i t} g(e ^{it}) \ d t \ ,
\EQNOD\defDofEXP
$$

\noindent where we used $dz=ie^{i\theta} d \theta$.  Observe that since
there is no $z^{-1}$ in $g$, $D(z)$ is again a Laurent polynomial
satisfying \CONDITIONSreformul.
Now we set $f(z)={{1}\over {g(z)}}$.  We see that $f(z)$ is bounded on
$S^1$ so, by the lemma above, $T_f$ is then
the bounded inverse of $T_g$.  Now we define the operator $X(z)=- T_f
{{\partial}\over {\partial z}}$:
$C^{\infty}(S^1)\rightarrow C^{\infty}(S^1)$.
By exactly the same computation as the one leading up to \DOPdef\
we obtain that on
$C^{\infty} \subset L^2(S^1)$, $X(z)$ and $D(z)$ so defined satisfy
the Heisenberg commutation relation as well as \CONDITIONSreformul.

Our original setup has been formulated in
terms of the $\ell ^2$ space.  Now we would like to return to it.
The map which maps back $L^2(S^1)$ to $\ell ^2$ is given by the Fourier
series method:
$f\rightarrow \{{ \hat f}(n)\}_{n\in {\Bbb Z}}$, where ${\hat f}(n)$
is the $n$-th Fourier coefficient of $f$.
We recall that using
\XzONE\ we can
describe the coefficients $\beta_n$ as:

$$
\beta _n={\hat f}(n+1), \quad n \in {\Bbb Z} \ .
\EQNOD\FRAPPATbetas
$$

\noindent {\bf Example}: $g(z)=a+bz^{-2}, |a/b|\ne 1$.  The condition
$g(1)=1$ gives $a+b=1$.  Thus
$D(z)=az+(1-2a)+(a-1)z^{-1}, |a/(1-a)|\ne 1$.
We have two cases to consider depending on whether
$ |a/(1-a)|<1$ (Case 1) or $|a/(1-a)|>1$ (Case 2).
\smallskip
\noindent{\it Case 1:} $f(z)=z^2{{1}\over {az^2 +(1-a)}}$.

\noindent We observe that $f$ is analytic inside the unit circle $S^1$ and thus
its Fourier expansion coincides with its Taylor expansion around $z=0$.
We get

$$
{\hat f}(n)=\cases{{(-1)^{(n-2)/2}\over 1-a} ({a \over 1-a})^{(n-2)/2},
&if $n\ge 2$ and $n$ is even;\cr
 0,& otherwise.\cr}
\EQNOD\EXAMPLEf
$$

\noindent Thus

$$
\beta _n=\cases{{(-1)^{(n-1)/2}\over 1-a} ({a \over 1-a})^{(n-1)/2},
&if $n\ge 1$ and $n$ is odd;\cr
 0,& otherwise.\cr}
\EQNOD\EXAMPLEbetas
$$

\noindent{\it Case 2:} $f(z)={1 \over a+(1-a)z^{-2}}, |a/(1-a)|>1$.

\noindent We
observe that $f$ is analytic outside of the unit circle $S^1$ and thus
its Fourier expansion coincides with its Taylor expansion around
$z=\infty$.  We get:

$$
{\hat f}(n)=\cases{{(-1)^{n/2}\over a} ({1-a \over a})^{(-n)/2},&if $n\le 0$ and $n$ is even;\cr
 0,& otherwise.\cr}
\EQNOD\EXAMPLEtwof
$$

\noindent Consequently,

$$
\beta_n =\cases{{(-1)^{(n+1)/2}\over a} ({1-a \over a})^{(-n-1 )/2},
&if $n\le -1$ and $n$ is odd;\cr
 0,& otherwise.\cr}
\EQNOD\EXAMPLEtwobetas
$$

This example illustrates very well the fact that the optimal
discretization is special  even from the point of view of the Heisenberg
commutation relations. Indeed the case excluded from the example is
$a=1/2$ which is precisely the optimal case.  We can get, however,
some information about this case making $a$ approach $1/2$.  This allows us to
discuss the formal solution \betaNoneFINAL.
One way of interpreting this solution relies on a regularization of $D$
which essentially amounts to moving the zeros of $D^{\prime}$ off
the unit circle.  We now present a simple example of such a regularization.
First we set $a={{1}\over{2}}-\epsilon, \epsilon >0$.  We observe that
$|{{a}\over {(1-a)}}|<1$, so, we are dealing with the first case above.
Directly from \EXAMPLEbetas\ we get:
$$
\beta _n(\epsilon)=
\cases{{(-1)^{(n-1)/2}\over 1/2-\epsilon} ({{1/2-\epsilon}\over
{1/2+\epsilon }})^{(n-1)/2},
&if $n\ge 1$ and $n$ is odd;\cr
 0,& otherwise.\cr}
\EQNOD\EXAMPLEregularized
$$
It is interesting to note that
$$
\beta _n(0)=\cases{2(-1)^{(n-1)/2},
&if $n\ge 1$ and $n$ is odd;\cr
 0,& otherwise.\cr}
\EQNOD\EXAMPLEregularizedZERO
$$
This is a formal solution to \genBETAeq\ and \genBETAsol,  with no symmetry
condition imposed on $\beta$`s.  It has exactly the same status as the
symmetric solution \betaNoneFINAL.  Thus we see that the formal solutions
of Section 5 can be justified through an appropriate regularization.
In addition we have to properly interpret (4.2) (classical limit).  It makes sense to
consider this expression for $\epsilon \ne 0$. Then, as it is easy to check,
$$
\sum _{k=-\infty}^{\infty} \beta _k(\epsilon) =1.
\EQNOD\BETASUMregularized
$$
The same expression does not make sense for $\epsilon =0$, clearly
indicating that one cannot interchange $\lim _{\epsilon \to 0} $ with
$\sum$.

The final item we would like to discuss in this section is the
question of Hermicity or rather lack thereof.  In our formulation the
pair $D$ and $X$ may not necessarily consist of (even formally) Hermitian
or adjoint elements.  This is plain, for example, if one looks at the
pair $D=E-1$ and $X=xE^{-1}$ or even more general pairs found above.
One could, however, make this pair adjoint to each other, thus turning $D$
and $X$ into a pair consisting of a
creation and annihilation operators, by constructing a proper representation
space.  One possibility would be to consider a Fock representation, which
carry an, essentially, unique Hermitian form.  Thus, for the last example,
the completion of
the space $span\{e_n=X^n\,1,n\ge 0\}$ with respect to the inner product
$<e_m,e_n> = \delta _{m,n}n!$
gives us a Fock space on which $D$ acts as an annihilation operator and $X$ acts as a
creation operator respectively, and both act as difference operators.  Another possibility is to consider a coherent representation, i.e. such which is
generated from the vector $v:\, Dv=\lambda v,\, \lambda \in \C$, whose special case is the Fock representation obtained for $\lambda=0$.  In the case of $D=E-1$ the vacuum state $v$ is a quasiperiodic function satisfying $v(x+1)=(\lambda +1)
v(x)$.  One obtains the representation space as the
$span\{e_n=X^n\,v,n\ge 0\}$ and the unique Hermitian form is defined
by $<v,v>=1, <e_m,e_n>=<v,D^mX^nv>$ where in the last expression one moves
$X$ to the left and $D$ to the right making use of \DzCOMMUTATOR, the formula
$<u,X^n w>=<D^nu,w>$ and $Dv=\lambda v$.

In the next section we study the optimal discretization for which we show that one can construct pairs of Hermitian (self-adjoint) operators $D$ and $X$.

\vskip\baselineskip
\vskip\baselineskip
\noindent{\bf\Chapter. MOMENTUM FORMULATION APPROACH TO OPTIMAL

  \  DISCRETIZATION }

\vskip\baselineskip

Now we take on the case of the optimal discretization.  Thus $M=N$
throughout this section.  Also, in this section we are interested in
self-adjoint pairs $X,P$ relative to the standard inner product on
$L^2(S^1)$.
For convenience we multiply $D$ from previous sections by $1/i$ and denote the resulting function by $D_N$.

The problem of classifying the pairs of operators satisfying Heisenberg
commutation relations under an additional assumption that one of the
operators be bounded was considered in [\REF\DORFMEISTER].  Since our
difference operators are bounded, we are studying a special case of
that classification, included in [\DORFMEISTER], in particular
in Theorem 8.5 therein.
The approach below
is direct, however, and can be used as an introduction to a more
encompassing
treatment of [\DORFMEISTER]. Moreover, we have a different
physical motivation, our interest lies in concrete bounded operators
explicitly given by
difference operators, reflecting the underlying assumption that the
coordinate
space is discrete. We recall some basic definitions from [\DORFMEISTER].
  Let $H$ denote a separable Hilbert space over $\Bbb C$.
\smallskip
\noindent{\bf Definition:} {\it Let $P$, $Q$ be operators in $H$, and
$\Omega$ a dense subspace of $H$.  We call $(P,Q)$ a conjugate pair on
$\Omega$ iff
\item{(K1)} $P$ and $Q$ are symmetric, $\Omega \subset D(P)\cap D(Q)$},
\vskip0.2\baselineskip
\item{(K2)} $P\Omega \subset \Omega, \, Q\Omega \subset \Omega$,
\vskip0.2\baselineskip
\item{(K3)} $P=\overline{P/\Omega}, \, Q=\overline {Q/\Omega}$,
\vskip0.2\baselineskip
\item{(K4)} $Q$ is bounded.
\smallskip
\noindent In this section we essentially show that our $X$ and $D_N$ form a conjugate
pair in the above sense.
\medskip
We start with
\smallskip
\noindent{\bf Lemma 2:} {\it Let $D_N(z)$ be optimal, then
$D^{\prime}_N(z)$ has two simple roots on the circle $S^1$}.
\smallskip
\noindent \Proof: \
Since $z=e^{i\theta}$, $0 \le \theta <  2\pi$,
we want to show that
${d\over d\theta} D_N \equiv D_N^\prime(\theta)$ has exactly two zeros for
$\theta \in [0, 2\pi)$.

\noindent To derive $D_N^\prime(\theta)$ we first compute $D_N(\theta)$. We
get that

$$
D_N(\theta) \ = \ 2  \ \sum_{k=1}^N {(-1)^{k+1} \over k}
{ (N!)^2 \over (N+k)! (N-k)!} \ \sin k\theta
\ .
$$

\noindent Hence,

$$
D_N^\prime(\theta) \ = \ 2  \ \sum_{k=1}^N (-1)^{k+1}
{ (N!)^2 \over (N+k)! (N-k)!} \ \cos k\theta
\
$$

\noindent or, in terms of $z$:

$$
\eqalign{
D_N^\prime(\theta) &=  \sum_{k=-N\atop k\ne0}^N
{ (-1)^{k+1}  (N!)^2 \over (N+k)! (N-k)!}  \ z^k = \cr
&= { (-1)^{N+1} (N!)^2 \over (2N)! } \ z^{-N}
\sum_{k=-N\atop k\ne0}^N (-1)^k  \pmatrix{2N \cr N \cr} z^k = \cr
&= { (-1)^{N+1} (N!)^2 \over (2N)! } \ z^{-N}
\left[ (1-z)^{2N} - (-1)^N
\pmatrix{2N \cr N \cr} z^N \right]  \ .  \cr
}
$$

\noindent Thus it suffices to show that the polynomial
$(1-z)^{2N} - (-1)^N \pmatrix{2N\cr N\cr} z^N$
has exactly two zeros for $z \in S^1$. Indeed,
we obtain:
$\left[ {(1-z)^2\over z} \right]^N = (-1)^N
\pmatrix{ 2N\cr N \cr} $
\ or \
$\sin^2{\theta\over2} = {1\over4} \root N \of{\pmatrix{2N\cr N\cr}}$.
This equation has two solutions, provided: \
${1\over4} \root N \of{\pmatrix{2N\cr N\cr}} < 1 \ , \quad N \ge 1$.

Indeed, for $N=1$ we get ${1\over4}\cdot 2 < 1$ and there are two
roots, $\theta={\pi\over2}, \ {3\over2} \pi$.
Now, we proceed by induction.
First we observe that the above inequality is equivalent to:
$\pmatrix{2N\cr N\cr} < 4^N$. We already proved it for $N=1$.
We note that:

$$
\pmatrix{ 2N+2 \cr N+1 \cr} = \pmatrix{ 2N \cr N \cr}
{(2N+1) (2N+2) \over (N+1)^2} =
\pmatrix{ 2N \cr N \cr} {2(2N+1)\over (N+1)}
\ ,
$$

\noindent and we use the induction hypothesis to prove the claim.

 To prove that the zeros are simple we show that
${d\over dz} [ (1-z)^{2N} - (-1)^N \pmatrix{ 2N \cr N \cr}
z^N ]$ is not zero there. Using that
$(1-z)^{2N} - (-1)^N \pmatrix{ 2N \cr N \cr} z^N = 0$
we obtain that this derivative can be zero if and only if
$2z^2-2z-1=0$. Since this equation has two real zeros not equal to
$\pm 1$ we conclude that the zeros of $D_N^\prime(\theta)$ are
simple.  \blackbox

 Now, we turn to the problem of the localization of those zeros.
We will show that both zeros move towards $\theta=\pi$ as
$N \rightarrow \infty$.
First we prove three technical lemmas.

\noindent{\bf Lemma 3:}

$$
{ (1 + {1\over 2N}) \over (1 + {1\over N} ) } \ < \
{ 1 + {1 \over 2(N+1)} \over 1 + {1 \over N+1} }
\ , \qquad N \ge 1 \ .
$$

\noindent{\bf Proof:}
Consider
$(1+{1\over 2N})(1+{1\over N+1}) = { 2N^2+5N+2\over 2N(N+1)}$
and
$(1+{1\over N}) (1+{1\over 2(N+1)}) = {2N^2+5N+3\over 2N(N+1)}$.
Hence,
$(1+{1\over 2N}) (1+ {1\over N+1}) < (1+{1\over N}) (1+{1\over 2(N+1)})$
which implies the claim. \blackbox

 The next lemma improves on the estimate used in the course of the proof
of Lemma 2.

\noindent{\bf Lemma 4:}

$$
\pmatrix{2N\cr N\cr} < 4^N \left[
{ (1+{1\over2N}) \over (1+{1\over N}) } \right]^N
\ , \qquad N \ge 1 \ .
$$

\noindent{\bf Proof:} The proof goes by induction on $N$. The only
nontrivial step is to observe that:
$\pmatrix{ 2(N+1) \cr N+1 \cr} = \pmatrix{ 2N \cr N \cr}
{4 (1+{1\over 2N}) \over (1+{1\over N}) }$
which, by the induction hypotheses, implies:

\noindent $\pmatrix{ 2(N+1)\cr N+1\cr} < 4^{N+1} \left[
{(1+{1\over 2N}) \over (1+{1\over N}) } \right]^{N+1}$.

\noindent Now, the claim follows from Lemma 3. \blackbox

 Finally, we have:

\noindent{\bf Lemma 5:} {\it The sequence
$a_N = {1\over 4} \root N \of{\pmatrix{2N\cr N\cr}}$ , \
$N=1,2,\ldots$ is increasing and

\noindent $\lim_{N\rightarrow\infty} a_N = 1$.
}

\noindent{\bf Proof:} By Lemma 4,
$a_N < {1+ {1\over 2N} \over 1+ {1\over N} }$.
Hence,

$$
4 \pmatrix{2N\cr N\cr} a_N < 4 \pmatrix{2N\cr N\cr}
{ (1+{1\over2N}) \over (1+ {1\over N}) }
\ .
$$

\noindent The left hand side equals
$\pmatrix{2N\cr N\cr}^{N+1\over N}$,
from which we get
$\pmatrix{2N\cr N\cr}^{N+1\over N} < \pmatrix{ 2(N+1)\cr (N+1)\cr}$,
which implies that: $a_N < a_{N+1}$.
To compute $\lim_{N\rightarrow\infty} a_N$ we observe that

\noindent $\lim_{N\to\infty} \exp {1\over N} \left[\ln(2N)! -
2 \ln (N)! \right]$
can be computed using the Stirling formula:

$$
\Gamma(x) \ = \
e^{-x} \ x^{x-{1\over 2}} (2\pi)^{1/2}
\left[ 1 + {1\over 12x} + \cdots \right]
\ ,
$$

\noindent valid for large $x$. Thus

$$\eqalignno{
\lim_{N\to\infty} a_N &= {1\over 4} \lim_{N\to\infty}
\exp {1\over N} \left[ \ln (2N+1)^{2N+{1\over2}} - 2 \ln
(N+1)^{N+{1\over 2}} \right] = \cr
&= {1\over 4} \lim_{N\to\infty} e^{ 2\ln \left( {2N+1\over N+1}
\right) }  = 1 \ , \cr
}
$$

\noindent thus completing the proof. \blackbox

 To find the roots of $D_N^\prime(\theta)$ we solve
$ \sin^2 {\theta_N\over 2} = a_N$.
By Lemma 5 we see that $\theta_N$'s are approaching $\pi$ as $N$
increases. Table 2 illustrates this phenomenon numerically.

\midinsert
\vskip\baselineskip
\noindent {\it Table 2. Roots of $D_N^{\prime}$.}
$$
\vbox{\offinterlineskip
\hrule height1pt
\halign{&\vrule width1pt#&\strut\quad\hfil#\quad
&\vrule width1pt#&\quad\hfil#\quad
&\vrule#&\quad\hfil#\quad
&\vrule#&\quad\hfil#\quad
&\vrule#&\quad\hfil#\quad
&\vrule#&\quad\hfil#\quad\cr
height3pt&\omit&&\omit&&\omit&&\omit&&\omit&&\omit&\cr
&\hfill N \hfill&&\hfill 1 \hfill &&\hfill 10 \hfill&&\hfill 100\hfill
&&\hfill 2~500 \hfill&&\hfill 10~000 \hfill &\cr
height3pt&\omit&&\omit&&\omit&&\omit&&\omit&&\omit&\cr
\noalign{\hrule height0.5pt}
height3pt&\omit&&\omit&&\omit&&\omit&&\omit&&\omit&\cr
& \hfill root 1  \hfill &&\hfill${1\over2}\pi$\hfill&&\hfill $2.32$ \hfill
&&\hfill$2.80 $\hfill &&\hfill$3.06 $\hfill &&\hfill$3.10 $\hfill &\cr
height3pt&\omit&&\omit&&\omit&&\omit&&\omit&&\omit&\cr
& \hfill root 2  \hfill &&\hfill${3\over2} \pi$\hfill &&\hfill$3.96 $\hfill
&&\hfill$3.48 $\hfill &&\hfill$3.27 $\hfill &&\hfill$3.19$\hfill &\cr
height3pt&\omit&&\omit&&\omit&&\omit&&\omit&&\omit&\cr
\noalign{\hrule height1pt}}
}
$$
\endinsert

 Let us denote by $\theta_N^+$, $\theta_N^-$ the zeros of
$D_N^\prime(\theta)$, $\theta^{\pm}_N \in [0, 2\pi ]$.
Then on the intervals $I^1_N = ( \theta_N^- - 2\pi, \theta_N^+)$,
$I^2_N = ( \theta_N^+, \theta_N^-)$,
$D_N(\theta)$ is monotone and thus $D_N(\theta)$ has an
inverse which
we denote by $\phi^1_N$ and $\phi_N^2$ respectively.
Now, we observe that the Hilbert space
$\H = L^2(S^1) \simeq L^2([\theta^-_N-2\pi, \theta^-_N])$
admits the direct sum decomposition:
$ \H = L^2(I^1_N) \oplus L^2(I_N^2) $.
Moreover, we have:

$$
L^2(I^i_N) \simeq L^2(D_N(I^i_N)) \ , \quad
\tau_i: \ \psi(\theta) \rightarrow \vert(D_N^\prime \circ \phi_N^i )
(y) \vert^{-1/2} \ \psi( \phi_N^i(y) ) \ ,
$$

\noindent where $y = D_N(\theta)$ and $i=1, 2$.
Now we show that $\tau$ is a unitary isomorphism.

\noindent{\bf Lemma 6:} {\it  $\tau_i$ is a unitary isomorphism, $i=1,2$. }

\noindent{\bf Proof:}

$$
\langle \psi_1 \vert \psi_2 \rangle =
\int_{I^i_N} \overline{\psi_1(\theta)} \psi_2(\theta) \ d\theta =
\int_{D_N(I^i_N)} \overline{\psi_1(\phi_N^i(y))}
\psi_2(\phi_N^i(y)) \ \Big\vert {dy\over d\theta}
\Big\vert^{-1} dy
\ .
$$

\noindent Observe, however, that $\Big\vert{dy\over d\theta}\Big\vert
= \vert (D_N^\prime \circ \phi^i_N) (y) \vert$. Hence,

$$
\langle \psi_1 \vert \psi_2 \rangle_{L^2(I^i_N)} \ = \
\int_{D_N(I^i_N)} (\overline{\tau_i \psi_1})(y) \ (\tau_i\psi_2)(y) \ dy
\ = \ \langle \tau_i \psi_1 \vert \tau_i \psi_2
\rangle_{L^2(D_N(I^i_N))}
\ ,
$$

\noindent that ends the proof. \blackbox

 We will need an explicit form of the inverse of $\tau_i$. A simple computation yields:

$$
\tau_i^{-1}: \ \phi \rightarrow
\vert D_N^\prime(\theta)\vert^{1/2} \phi( D_N(\theta) ) \ ,
\quad \theta \in I^i_N \ , \quad \phi \in L^2(D_N(I^i_N))  \ .
$$

\noindent Now, we want to define the skew-Hermitian part of the
operator ${1\over D_N^\prime(\theta)}
{d\over d\theta} {1\over D_N^\prime(\theta)}$ on $L^2(D_N)$.
It is easier, however, to consider first its push-forward under
$\tau$:

$$
\eqalign{
&\ \ {1\over 2} \tau_i \circ \left[ {1\over D_N^\prime(\theta)}
{d\over d\theta} + {d\over d\theta} {1\over D_N^\prime(\theta)}
\right] \circ \tau_i^{-1} \phi(y) = \cr
&= {1\over 2}\tau_i \circ \left[ {1\over D_N^\prime(\theta)}
{d\over d\theta} + {d\over d\theta} {1\over D_N^\prime(\theta)}
\right]
(\sign(D_N^\prime(\theta)) \ D_N^\prime(\theta))^{1/2}
\ \phi(D_N(\theta)) = \cr
&= {1\over2} \tau_i \circ \left[
{1\over D_N^\prime(\theta)} \ \phi(D_N(\theta)) +
2 {\sign D_N^\prime(\theta)\over ( \sign( D_N^\prime(\theta) )
D_N^\prime(\theta) )^{1/2} } \ \partial_\theta \phi(D_N(\theta))
\right] = \cr
&= \tau_i \circ
{\sign D_N^\prime(\theta) \over (\sign( D_N^\prime(\theta) )
D_N^\prime(\theta) )^{1/2} } \
\partial_\theta \phi(D_N(\theta)) = \cr
&= \vert D_N^\prime \circ \varphi^i_N(y)\vert^{-1/2} \
{ \sign(D_N^\prime(\theta)) \circ \varphi^i_N)(y) \over
\vert D_N^\prime(\theta) \circ \varphi_N^i(y)\vert^{1/2} }
{\partial\over\partial\varphi^i_N(y)} \phi(y) = \cr
&= {1\over D_N^\prime \circ \varphi_N^i) (y)}
{\partial \over \partial \varphi^i_N(y) } \phi(y)
\ .  \cr
}
$$

\noindent Since ${\partial\over\partial y} =
{\partial\varphi_N^i(y)\over\partial y} {\partial \over\partial
\varphi_N^i(y)}$ we need to compute ${\partial \varphi_N^i(y)\over\partial y}$.  To this end  we simply
observe that $(\varphi_N^i\circ D_N) (\theta) = \theta$. Hence,
$$
{d\varphi_N^i\over dy} = {1\over D_N^\prime(\theta)} =
{ 1\over (D_N^\prime \circ \varphi_N^i) (y)}
\ .
$$

\noindent This implies:

$$
\left( \tau_1 \circ {1\over2} \left[
{1\over D_N^\prime(\theta)}
{d\over d\theta} + {d\over d\theta} {1\over D_N^\prime(\theta)}
\right] \circ \tau_1^{-1} \right) (y) =
{\partial\over\partial y} \ ,
\quad y\in D_N(I_N^i) \ .
$$

\noindent We thus have the following important

\vskip0.4ex
\noindent{\bf Theorem 4:} {\it
Let $X_N = {i\over2} \left[
{1\over D_N^\prime(\theta)} {d\over d\theta} +
{d\over d\theta} {1\over D_N^\prime(\theta)}
\right]$
be defined on a dense set $S\in L^2(S^1)$.  Then $X$ is unitary equivalent to
$i{\partial\over\partial y}\Big\vert_{L^2(D_N(I_N^1))}
\oplus
i{\partial\over\partial y}\Big\vert_{L^2(D_N(I_N^2))} $, defined on
the dense space $\tau _1(S\cap L^2(I_N^1))+\tau _2(S\cap L^2(I_N^2))$.
}
\vskip0.4ex

\noindent We can prove a similar statement for $D_N(\theta)$.
We have:
$\tau_i\circ D_N(\theta)\circ\tau_i^{-1} \phi(y) =
\tau_1\circ D_N(\theta) \vert D_N^\prime(\theta)\vert^{1/2}
\phi(D_N(\theta)) $
$= \vert D_N^\prime \circ \varphi^i(y)\vert^{-1/2} D_N\circ\varphi^i(y)
\vert D_N^\prime \circ \varphi^i(y) \vert^{1/2}
\phi(D_N\circ\varphi^i)(y)$
$= y \phi(y)$ , \  $y \in D_N(I^i_N)$.
Hence

\vskip0.4ex
\noindent{\bf Theorem 5:}  {\it
Let $P_N$ be the operator of multiplication by the function $D_N$.  Then
$P_N$ is  unitary equivalent to
$y\Big\vert_{L^2(D_N(I^1_N))} \oplus y\Big\vert_{L^2(D_N(I^2_N))}$.
}
\vskip0.4ex

\noindent{\bf Corollary:} The pair $Q=D_N$ and $P=-X$ is a conjugate
pair on $\Omega$, where:
$\Omega=\tau ^{-1}(S_1)\oplus \tau^{-1}(S_2)$,
$S_{i}= \{ f \in L^2(D_N(I_N^i)), f$
is absolutely continuous,  $f=0$
on the boundary of $D_N \},$   $i=1,2$.

\vskip0.4ex
\noindent {\bf Remark} Observe that $X$ is closed and symmetric on
$\Omega$.
We discuss its self-adjoin extensions below.

First, however, we determine the spectrum of $D_N$ or to be more precise
$T_{D_N}$.
By the well known theorem on Toeplitz operators
[\REF\WIDOM]: Spec$(T_{D_N}) =$ essential range of $D_N =$
range of $D_N$, where the latter is a consequence of continuity.
Because $X_N$ decomposes into a direct sum of operators of the type
$i {d\over dy}$ on the finite intervals $D_N(I^i_N)$ we present
a brief description of basic features of such operators.
For convenience
we take a finite interval $[-1, 1]$. The following discussion is based
on [\REF\REEDSIMON] and all the details can be found there.
The operator $i {d\over dy}$ is known to have a one parameter family of
self--adjoint extensions. More precisely, let us set
$i {d\over dy} = T_\alpha$, defined on
$D(T_\alpha) = \left\{ \varphi: \varphi \right.$
is absolutely continuous and
$\left. \varphi(-1) = \alpha \varphi(1) \right\}$, where
$\alpha \in \C, \ \vert\alpha\vert = 1$.

 To interpret $\alpha$ we solve the eigenvalue problem:
$i {d\over dy} \psi = \lambda \psi$ with
$\psi(-1) = \alpha \psi(1) = e^{i\varphi_0} \psi(1)$,
$\alpha = e^{i\varphi_0}, \ \varphi_0 \in [0, 2\pi)$.
Thus we get: $\psi(y) = e^{i\lambda y}$ and
$e^{-i\lambda} = e^{i\varphi_0} e^{i\lambda}$
from which it follows that $2\lambda + \varphi_0 = 2 \pi n$, $n\in \Z$.
Hence, $\lambda = \lambda_n = n\pi - {\varphi_0\over 2}$, $n\in Z$.
The appearance of $\pi$ is accidental and can easily be removed
by rescaling of $T_\alpha$. The spectrum of $T_\alpha$
is a lattice of physical positions. We would like to point out that
there is no canonical way of identifying this lattice with the lattice
we have started with. It seems to be compelling in fact to consider as
a real physical space the lattice of eigenvalues of $T _{\alpha}$
and the corresponding Hilbert space $L^2(D_N(I^1_N)$, for example.  Thus
we would, so to speak, avoid the ``spectrum doubling `` problem by decree.
On the other hand this would have to be interpreted quantum mechanically as
saying that the process of quantization amounts in this case to finding
an irreducible representation of the Heisenberg commutation relations with
the additional condition that the spectrum of would be position is discrete.
This representation would then have a relation to the calculus of difference
operators as described above, yet, the physical Hilbert space would be only
``half'' of the Hilbert space that naturally carries the
action of difference operators.

To close this section we would like to mention that the optimal discretization
has a very nice property that $\lim_{N \to  \infty}D_N(\theta)=\theta$.
This fact is proven in Appendix~B. It is only in this case that we get
an irreducible representation of the Heisenberg commutation relations in the
case of the optimal discretization.

%%%%%==========================================================

\vskip\baselineskip
\vskip\baselineskip
\noindent{\bf\Chapter. SUMMARY AND CONCLUSIONS}
\vskip\baselineskip

 In this paper we have investigated a wide class of discretization
schemes for the derivative and/or momentum operator defined by
\DERIVorig.
The optimal subclass of these schemes has been distinguished, that
give the best fit to the continuous operators and are symmetric
with respect to the space reflections.
The particular form of those schemes has been determined in Sec.~2.
It has been shown that the solution found in [\TURBINERSMIR]
is the only solution in the form of finite series for the operator $X$.
We also have determined that the commutation relation for the
annihilation--creation type of operators can be satisfied for a large
class of difference operators, thus extending the result of
[\TURBINERSMIR], if one admits $X$ in a form of an infinite series
in $E$.

 Our analysis in Sec.~6 has shown that upon quantization the
classically different optimal
discretization schemes are all unitary equivalent.  As a result,
the number of points used to ``delocalize'' ${d}\over {dx}$ is
unessential.  For a fixed number of these points, say $2N+1$,
the representation of the canonical commutation relations is reducible.
Furthermore, by the result of Appendix~B its irreducible components
are unitary equivalent to the case with an infinite
($N\rightarrow \infty$) number of points.

 Further extensions of this work can be done in several directions.
First, one can investigate the case of an imaginary shift, as has
been mentioned in Sec.~1.
Second, more general Ansatz for the $P$ and $X$ operators can be
analyzed. In particular, we suggest that the representation used
in [\DOUGLAS] can be included into our scheme by considering operator
$X$ being a nonlinear function of $x$. Finally, it would be interesting
to study simple quantum mechanical systems within our approach.

\vskip\baselineskip
\noindent{\bf Acknowledgment.} We are greatly indebted to P.O.~Mazur
for his insightful comments and constant encouragement.
Also, we thank A.~Herdegen for many illuminating
suggestions regarding the content of Sec.~6.  The authors gratefully
acknowledge the support of the State Committee for Scientific Research
Republic Poland (KBN) 2 P03B 140 10
and the Natural Sciences and Engineering Research Council of Canada.
AZG would like to thank the Ko\'sciuszko Foundation (NY) for the
financial support and the University of South Carolina
for hospitality.

%%%\vfill\eject
\vskip\baselineskip
\vskip\baselineskip
\vskip\baselineskip
\centerline{\bf A~p~p~e~n~d~i~c~e~s}

\Chapterstart1

\vskip\baselineskip
\vskip\baselineskip
\noindent{\bf Appendix A:  Heine symbols}
\vskip\baselineskip

The Heine symbol defined by \HEINE\
have pretty interesting behavior.
The deformation parameter $q\ge 0$ is usually bound to the (complex)
region: $\vert q\vert \le 1$ and it is parameterized exponentially by a
(dimensionless) physical parameter $l$:

$$
q = \exp \left( {i l \over 2}  \right)
\ . \eqno(A.1)
$$

\noindent The above parametrization allows us to rewrite the Heine
symbol \HEINE\ in the following widely used form:

$$
[\, P \, ]_q \ \equiv \ { \sin({l P\over2}) \over \sin({l\over2}) }
\ = \ { \sinh({i l P\over2}) \over \sinh({i l\over2}) }
\ , \eqno(A.2)
$$

Function $H(P,q)$ is antisymmetric in $P$:

$$
H(P,q) \ = \ - H(-P,q)
\ . \eqno(A.3)
$$

\noindent Also, it has the following limits for fixed $q$:

$$
\eqalign{
&\lim_{P\to 0} H(P,q) = 0 \ , \cr
&\lim_{P\to 1}H(P,q) = 1 \ , \cr
&\lim_{P\to \infty} H(P,q) = \infty \quad(q \ne 0, 1) \ . \cr
} \eqno(A.4)        $$

 As a function of $q$ the function $H(P,q)$ is invariant under
the transformation:

$$
q \rightarrow q' \equiv {1\over q}
\ . \eqno(A.5)
$$

\noindent For $P=1$ it is constant: $H(1,q) = 1$ and for $P < 1$
($P > 1$) it has maximum (minimum) at $q = 1$ equal to $P$.
The point $P=q=1$ is singular and
the limits $\lim_{q\to 1}$ and $\lim_{P\to 1}$ do not commute.
Also, the limit $\lim_{q\to 0}$ is singular --- it is equal 0
for $P\in[0,1)$, equal 1 for $P=1$ and undefined for $P>1$.

\Chapterstart1

\vskip\baselineskip
\vskip\baselineskip
\noindent{\bf Appendix B:  A remark about $D_N$ }
\vskip\baselineskip

Let us consider the function $f(\theta)=\theta, \, -\pi < \theta < \pi$.
This function has the Fourier expansion:
$$
\theta=\sum _{n=1}^{\infty} b_n \sin n\theta,
$$
where
$$
b_n={2(-1)^n \over n}\, .
$$
Let us recall that
$$
D_N(\theta)=i\sum _{n=1}^N {2(-1)^{n+1}(N!)^2 \over n (N+n)! (N-n)!}\sin n
\theta.
$$
The main idea now is to compare the $N$-th partial sum $S_N$ of $if(\theta)$
 with the above sum.  Clearly,
$$
S_N(\theta)=i\sum _{n=1}^N {2(-1)^{n+1} \over n }\sin n\theta.
$$
The two sums look alike, the difference is the factor
$ {(N!)^2 \over (N-n)!(N+n)!}$.

We claim that $D_N(\theta)$ is an approximation to $i\theta$, i.e.
$\lim _{N \to \infty} D_N(\theta)=i\theta $ in $(-\pi, \pi)$.
This approximation is a result
of a specific method of summation of the Fourier series, similar to the
Fejer method of arithmetic means. Let us briefly review this subject.
We refer the reader to [\REF\Zygmunt] for more details.  Given an infinite
matrix $a=[a_{ij}: i,j=0,1, \dots]$ and a sequence $\{S_n: n=0,1, \dots\}$
we define a new sequence $\sigma _n, n=0,1, \dots$ by
$$
\sigma _N=\sum _{n=0}^{\infty}a_{N n}S_n.
$$
We require that $a$ satisfy the following conditions:

\vskip0.1\baselineskip
\noindent (i) $\lim _{N\to\infty} a_{N n}=0$.
\vskip0.3\baselineskip

\noindent (ii) $B_N:= \sum_{n=0}^{\infty} |a_{N n}|$
exists for every $N=0,1,\dots$,  and the set $\{B_N:N=0,1,\dots\}$ is bounded.
\vskip0.3\baselineskip

\noindent (iii)  $\lim _{N\to \infty} \sum_{n=0}^{\infty} a_{Nn} =1$.
\vskip0.3\baselineskip

\noindent If the $\sigma _N$ have a limit $s$ then we say that the sequence of partial
sums $S_N$ is $\it {a-summable}$ to the limit $s$. The most important fact for us
is the following:
\smallskip
\noindent {\bf Theorem B.1 [\Zygmunt]}:
{\it If $a$ satisfies (i), (ii) and (iii) and if ${S_N}$ tends to
a finite limit $s$, then $\lim_{N \to \infty}\sigma _N =s$.}
\smallskip
\noindent We claim that $D_N$ are precisely the linear
means $\sigma _N$ for an
appropriate choice of $a$.  We define:
$$
a_{Nn} =\cases{{{(2n+1)(N!)^2}\over {(N-n)!(N+n+1)!}}\,,
&\quad if $0<n\le N$ ;\cr
 0,& \qquad otherwise.\cr}
$$
It is easy to check that $a$ so defined satisfy (i) and (ii). To see that
(iii) is satisfied as well we observe that
$a_Nn$ satisfy:
$$
\sum _{m=n}^N a_{Nm}={{(N!)^2}\over {(N-n)!(N+n)!}}, \qquad
0 \le n \le N.\eqno (B.1)
$$
In particular, for $n=1$ we obtain:
$$
\sum _{n=1}^N a_{Nn}={{(N!)^2}\over {(N-1)!(N+1)!}}={{N}\over {N+1}}.
$$
By taking the limit $N\to \infty$ we obtain (iii).  To prove our claim we
need to show that $D_N=\sigma _N$.  This follows from $(B.1)$.  Finally,
applying Theorem B.1 we conclude that
$\lim _{ N \to \infty}D_N(\theta)=i\theta$.

\vskip3\baselineskip

%   REFERENCES
%  ============

\noindent REFERENCES
\vskip0.5\baselineskip

%--------------------------------------------------------------------

\item{[\AMBARCUMIAN]}     \AMBARCUMIANref
\item{[\RIEMANN]}         \RIEMANNref
\item{[\GIBBS]}           \GIBBSref
\item{[\KLEBANOV]}        \KLEBANOVref
\item{[\PENROSEspincalc]} \PENROSEspincalcref
\item{[\LOOPREPR]}        \LOOPREPRref
\item{[\SMOLIN]}          \SMOLINref
\item{[\BEKENSTEIN]}      \BEKENSTEINref
\item{[\MUKHANOV]}        \MUKHANOVref
\item{[\JACOBSON]}        \JACOBSONref
\item{[\BEKENSTEINMUKHANOV]} \BEKENSTEINMUKHANOVref
\item{[\PABLOAPPb]}       \PABLOAPPbref
\item{[\THOOFTCA]}        \THOOFTCAref
\item{[\THOOFTa]}         \THOOFTaref
\item{[\THOOFTb]}         \THOOFTbref
\item{[\AHARONOVPETERSEN]} \AHARONOVPETERSENref
\item{[\WILSON]}          \WILSONref
\item{[\REGGE]}           \REGGEref
\item{[\SORKIN]}          \SORKINref
\item{[\GARAY]}           \GARAYref
%\item{[\BOPP]}            \BOPPref
%\item{[\BORN]}            \BORNref
%\item{[\PAIS]}            \PAISref
%\item{[\CHODOS]}          \CHODOSref
\item{[\LUKIERPoincare]}  \LUKIERPoincareref
\item{[\PABLOAPPa]}       \PABLOAPParef
\item{[\PABLOPRL]}        \PABLOPRLref
\item{[\DOUGLAS]}         \DOUGLASref
\item{[\DIRACQUANT]}      \DIRACQUANTref
\item{[\TURBINERSMIR]}    \TURBINERSMIRref
\item{[\COUTINHO]}        \COUTINHOref
\item{[\COLE]}            \COLEref
\item{[\FRAPPAT]}         \FRAPPATref
\item{[\DORFMEISTER]}      \DORFMEISTERref
\item{[\WIDOM]}           \WIDOMref
\item{[\REEDSIMON]}       \REEDSIMONref
\item{[\Zygmunt]}         \Zygmuntref

\bye